\documentclass[journal]{IEEEtran}
\usepackage{amsmath,epsfig,amssymb,verbatim,amsopn,subfigure,color}
\usepackage{hyperref} 
\usepackage{cite,xspace}
\usepackage{array,algorithm,algorithmic}
\usepackage{bbm}
\usepackage{array}
\usepackage{multirow}
\usepackage{multicol}
\usepackage{rotating}
\usepackage{dsfont,float}
\usepackage{stfloats}
\usepackage{enumerate,makecell}
\usepackage{graphicx}
\usepackage{graphbox}
\usepackage{cases}
\usepackage{float}
\normalsize

\makeatletter
\let\l@ENGLISH\l@english
\makeatother
\makeatletter
\renewcommand*{\@opargbegintheorem}[3]{\trivlist
  \item[\hskip \labelsep{\itshape #1\ #2}] {\itshape (#3):} {\normalfont}}
\makeatother
\usepackage{relsize}

\graphicspath{{figs/},{Figures/}}

\author{IEEE Publication Technology,~\IEEEmembership{Staff,~IEEE,}
\thanks{This paper was produced by the IEEE Publication Technology Group. They are in Piscataway, NJ.}
\thanks{Manuscript received xxx; revised xxx.}}
\newcommand{\AuthorOne}{Weihao Wang}
\newcommand{\AuthorTwo}{Jing Guo, {\em{Senior Member, IEEE}}}
\newcommand{\AuthorThree}{Siqiang Wang}
\newcommand{\AuthorFour}{Xinyi Wang, {\em{Member, IEEE}}}
\newcommand{\AuthorFive}{Weijie Yuan, {\em{Senior Member, IEEE}}}
\newcommand{\AuthorSix}{Zesong Fei, {\em{Senior Member, IEEE}}}
\newcommand{\ThankOne}{W. Wang, J. Guo, S. Wang, X. Wang, and Z. Fei are with the School of Information and Electronics, Beijing Institute of Technology, Beijing 100081, China (Emails: \{weihaowang, jingguo, 3120205406, wangxinyi, feizesong\}@bit.edu.cn).

Weijie Yuan is with the School of Automation and Intelligent Manufacturing, Southern University of Science and Technology, Shenzhen 518055, China (e-mail: yuanwj@sustech.edu.cn). 
}

\begin{document}

\title{Pre-equalization Design for ISAC-OTFS Air-Ground Transmission: A Deep Learning Approach}
\author{\IEEEauthorblockN{\AuthorOne,~\AuthorTwo,~\AuthorThree,~\AuthorFour,\\~\AuthorFive~and~\AuthorSix\thanks\ThankOne}}

\maketitle

\begin{abstract}
Despite the strong Doppler resilience capability, orthogonal time-frequency space (OTFS) modulation suffers from high channel estimation and equalization complexity at the receiver, hindering its applicability in air-ground transmission. In this paper, we propose a pre-equalization–based integrated sensing and communications-OTFS downlink transmission framework in which the terrestrial access point executes pre-equalization using the predicted channel state information (CSI), so that the unmanned aerial vehicle can perform direct symbol detection without channel equalization. In particular, the mean square error of OTFS symbol demodulation and Cramér-Rao lower bound of sensing parameter estimation are considered, with their weighted sum utilized as the metric for optimizing the pre-equalization matrix. To address the time-varying CSI, we develop a deep learning based framework composed of channel prediction and pre-equalization. In particular, a parameter-level channel prediction module is utilized to decouple OTFS channel parameters, and a low-dimensional prediction network is leveraged to correct outdated CSI, which is then used to initialize the input of the pre-equalization module. Finally, a dual-branch residual-structured deep neural network is cascaded to execute pre-equalization. Simulation results show that the proposed channel prediction-based pre-equalization framework significantly reduces receiver complexity and pilot overhead while achieving symbol detection performance close to minimum mean square error equalization with perfect CSI under high mobility, as well as substantially improving sensing accuracy.

\end{abstract}
%

\ifCLASSOPTIONpeerreview
    \newpage
\fi

\vspace{-0mm}\section{Introduction}\label{sec:intro}
Efficient development of the low-altitude economy relies on close collaboration of aerial and terrestrial infrastructures \cite{10879807}. As a result, air-ground transmission has attracted significant research attention in recent years \cite{10329470,jing2024isac}. The two most critical perspectives in air-ground networks are exceptional wireless communication and sensing capabilities, which enable data transmission, intrusion detection, and trajectory planning \cite{10879807,10168298,11207645}, etc. However, the limited spectrum resources hinder simultaneous high-throughput communication and high-accuracy sensing in the air-ground network, necessitating higher spectral efficiency techniques to meet this demand. The integrated sensing and communications (ISAC)
technology is regarded as a promising candidate \cite{liu2022integrated, liu2022survey,lu2024integrated,wei2023integrated}. ISAC technology utilizes a unified waveform to simultaneously achieve communication and sensing functions. This unified approach shares hardware resources, spectrum resources, and signal processing algorithms, leading to benefits such as improved spectrum and energy efficiency, and reduced hardware costs. 

It is worth noting that the air-ground integrated network can involve the high-mobility scenarios. In such cases, the wireless channels may transition from stable to highly dynamic, or from time-invariant channel to fast time-variant channel, degrading communication and sensing performance \cite{xiao2021overview}. For example, the performance of ISAC systems using multi-carrier waveforms (e.g., orthogonal frequency division multiplexing) can be severely degraded by inter-carrier interference \cite{shtaiwi2024orthogonal}. Hence, how to achieve efficient communication and precise sensing under high mobility has emerged as a critical challenge for air-ground ISAC networks.

Among the possible candidates, orthogonal time frequency space (OTFS) exhibits high reliability for communication and sensing compared to other commonly used waveforms \cite{wu2021otfs}. Besides, OTFS offers a unique advantage for ISAC applications since its communication channel parameters and sensing parameters share the same form. Hence, it has received extensive attention from researchers in recent years \cite{9903393,10638525,wang2023sensing}. For example, in \cite{zegrar2024otfs}, an OTFS-based ISAC scheme was proposed, allowing for highly accurate range-velocity profiles without requiring large bandwidth transmissions or long-duration frames. The authors in \cite{shi2023integrated} addressed the challenge of precise estimation of user state information alongside reliable data transmission in high-mobility environments. A deep learning based target detector was designed in \cite{suarez2023deep}, which takes the delay–Doppler correlation matrix as input and learns to recognize target signatures through training over multiple channel and noise scenarios, thereby achieving adaptive target detection without threshold design. Given that ISAC systems typically has to sense environmental parameters to perform specific tasks, some studies shifted their focus from mere target detection to parameter estimation. For example, in \cite{wu2024optimal,10791452}, the authors minimized the bit error rate (BER) of the OTFS-based ISAC system while ensuring accurate sensing parameter estimation. The authors in \cite{keskin2024integrated} proposed a generalized likelihood ratio test (GLRT) based multi-target detection and delay-Doppler-angle estimation algorithm for multiple-input-multiple-output OTFS radar sensing. Through jointly optimizing the spatial-delay-Doppler domain waveform, a better trade-off between sensing SNR and communication rate was achieved. 

Note that the aforementioned works were all based on perfect channel state information (CSI); however, in practical applications, the high mobility of the aerial nodes can result in significantly outdated CSI, which will definitely degrade the communication performance. To address this issue, some studies have attempted to predict the spatial domain channel to dynamically adjust beamforming. Specifically, the authors in \cite{yuan2023orthogonal} used the sensing echoes to assist in beam alignment and beam tracking, where Kalman filtering was employed to predict the angle-domain information for updating beamforming strategies. A convolutional long-short term memory recurrent neural network (CLRNet) was then proposed in \cite{liu2022deep} by introducing convolutional neural network modules for spatial feature extraction and long short-term memory (LSTM) modules for capturing temporal dependencies. In \cite{9492131}, based on deep learning, the roadside unit utilized vehicle angular parameters obtained from sensing echoes to predict beam directions for the next time instant, thereby achieving precise beam pairing with minimal latency. These works focused on channel prediction in the spatial domain, while there have been little works studying methods to obtain the delay and Doppler shift information of OTFS.

For delay-Doppler (DD) domain CSI, a novel sensing-assisted channel estimation algorithm was proposed in \cite{yuan2021integrated}, where the BS firstly senses the topology of propagation environment, and then reconstructs the communication channel using the sensing parameters. Specifically, the reconstructed topology was adopted to assist in predicting the channel parameters, including delay and Doppler shift, based on kinematic formulas, such that only the fading of each path has to be estimated via pilots. On the other hand, the authors in \cite{liu2023predictive} capitalized on past channel information across multiple time slots to forecast the next communication precoding matrix, resulting in lower frame error rate and achieving ultra-reliable low-latency communications. Although the aforementioned works enabled reduced pilot overhead, the significant channel equalization processing complexity of the OTFS receiver has not been addressed. This oversight may render them unsuitable for aerial receivers with limited signal processing capabilities. Besides, the sensing perspective was also not considered in the aforementioned works.

In this work, we explore the air-ground OTFS-based ISAC downlink transmission system. By jointly considering the outdated channel caused by high mobility and the limited signal processing capabilities at the aerial receiver, a channel prediction-based pre-equalization (CP-PE) framework is proposed to achieve a better trade-off between communication and sensing performance. The main contributions of this paper are concluded as follows:

\begin{itemize}
\item  For the air-ground network with one terrestrial access point (AP), one unmanned aerial vehicle (UAV) UE, and multiple scatters, an ISAC downlink transmission framework based on OTFS waveform is developed, which includes a pre-equalization module to reduce the complexity at the UE receiver side. Under this transmission framework, the mean square error (MSE) of OTFS data symbols and the Cramér-Rao lower bound (CRLB) of sensing parameter estimation are analyzed. Based on this, we formulate a weighted MSE and CRLB optimization problem under the transmission power constraint to balance the communication and sensing performance.

\item To address the formulated complicated optimization problem as well as outdated CSI, we propose a CP-PE framework by leveraging the nonlinear mapping capability of deep learning. Specifically, the CP-PE framework consists of three modules. The first is a parameter-level channel prediction module with a low-dimensional input space, which directly corrects outdated CSI by utilizing the decoupled OTFS channel parameters. The second is a CSI processing module that performs post-processing on the parameter-level channel data. The third is a dual-branch residual-structured deep neural network (DNN)-based pre-equalization module, which jointly considers both communication and sensing performance.

\item  Simulation results show that the proposed framework effectively enhances communication and sensing performance compared to pre-equalization design using outdated CSI and can approximate the performance achieved with true CSI. This indicates the robustness of the proposed framework for pre-equalization in high-mobility scenarios. Furthermore, the processing complexity at the UAV is reduced by more than 75\% compared to the conventional scheme.
\end{itemize}

The remainder of this paper is organized as follows. Section~\ref{sec:model} presents the system model, including the pre-equalization framework for ISAC signals. Section~\ref{sec:Problem} analyzes the performance metrics of the proposed transmission framework and then formulates the optimization problem, considering the weighted communication and sensing performance. The CP-PE framework, including the channel prediction module, CSI processing module, and pre-equalization design module, is elaborated in Section ~\ref{sec:algorithm}. In Section~\ref{sec:result}, numerical results are discussed to evaluate the communication and sensing performance of the proposed algorithms under different parameter settings. Finally, the paper is concluded in Section~\ref{sec:concl}.

This paper adopts the following notations. A boldface capital letter, a boldface lowercase letter, and a calligraphy letter are used to denote a vector, a matrix, and a set, respectively. $\mathbf{I}_{N}$ denotes the $N\times N$ identity matrix. The transpose and Hermitian transpose are denoted by $\left(\cdot\right)^T$ and $\left(\cdot\right)^H$, respectively. $\otimes$ denotes the Kronecker product. $\mathrm{vec}\left(\cdot\right)$ represents converting a matrix $\cdot$ to a vector column by column. $\mathbb{E}_{n}\left[\cdot\right]$ represents taking the average of $\cdot$ with $n$ as the variable. $\operatorname{diag}\{\cdot\}$ denotes a diagonal matrix. $\mathbf{F}_N$ represents the discrete Fourier transformation matrix of size $N\times N$. $\|\cdot\|_2$ denotes the 2-norm of its argument.

\vspace{-0mm}\section{System Model}\label{sec:model}
\begin{figure}[t]
\centering
\includegraphics[width=0.7\linewidth]{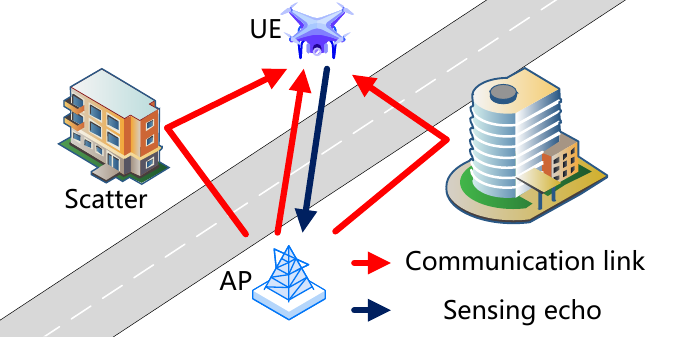}
\caption{An illustration of the air-ground ISAC system model.}
\label{fig:system model}
\end{figure}

We consider an ISAC-enabled air-ground downlink transmission system under the high-mobility communication scenario, which consists of one AP, one UAV UE, and several scatters, as shown in Fig. \ref{fig:system model}. Therein, the AP simultaneously senses and communicates with the UE via the OTFS waveform. As this study concentrates on pre-equalization design in the DD domain instead of spatial-domain, both the AP and UE are assumed to employ a single antenna, adopting the same setting as the studies in \cite{tao2024channel,zhang2023generalized,10329933}.

\subsection{ISAC Pre-equalization Framework}\label{subsec:framework}
To reduce spectrum resource occupancy and signal processing complexity of UE, we propose a pre-equalization framework for ISAC-OTFS downlink transmission systems. In particular, pre-equalization is a scheme that mitigates the impact of the channel at the transmitter (AP) side, thereby avoiding the complex channel estimation and equalization at the UE. Specifically, the comparison between the block diagrams of the pre-equalization based ISAC framework and conventional OTFS system is illustrated in Fig. \ref{fig:process}. Different from the conventional scheme, the OTFS modulated signal $\mathbf{s}_t$ is firstly pre-equalized by the pre-equalization matrix $\mathbf{P}_t$ at AP and then transmitted to UE, which directly performs symbol detection without channel estimation and channel equalization. This approach can significantly reduce the signal processing complexity at the receiver. For the sensing process, only the aerial UE echo is considered, as clutter can be effectively suppressed using conventional moving target indication and clutter mitigation techniques \cite{7862855,192088}. This means a unified waveform is required to simultaneously fulfill both communication and sensing functions. In this work, we primarily focus on pre-equalization design at AP to accommodate both communication and sensing. The following design principles are taken into account during pre-equalization design:

\subsubsection{Trade-off Between Communication and Sensing Performance}
There is a contradiction between the communication and sensing performance for a specific signal. Therefore, pre-equalization design needs to achieve a trade-off between communication and sensing performance under a specific requirement, thus improving the spectrum resource utilization.
\subsubsection{Outdated CSI and Sensing Parameters}
Note that waveform design generally requires the instantaneous CSI and sensing parameters, while in most practical applications, only outdated CSI and sensing parameters are available. Therefore, it is essential to design an algorithm that accounts for outdated CSI and sensing parameters to minimize the impact of pre-equalization design inaccuracies on communication and sensing performance.

\begin{figure*}
    \centering
    \includegraphics[width=0.95\linewidth]{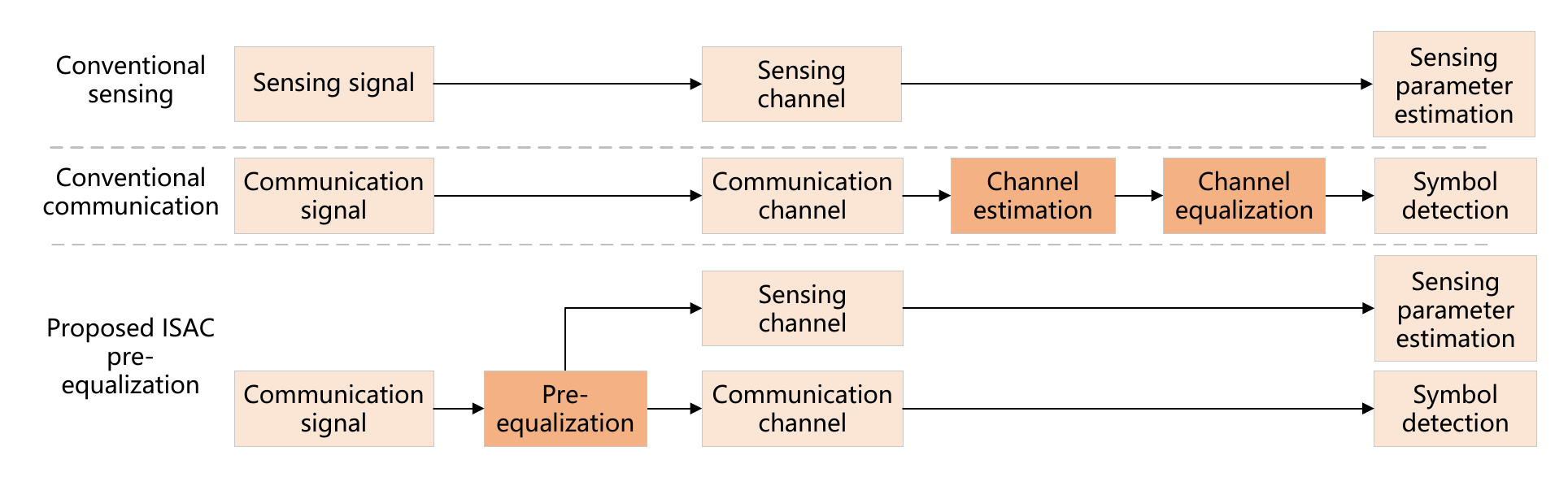}
    \caption{Comparison between the conventional framework and the proposed ISAC pre-equalization framework.}
    \label{fig:process}
\end{figure*}
\subsection{Sensing Model}
Let $\mathbf{s}_t$ denote the original OTFS modulated symbol in the DD domain. Therein, the modulated symbols in $\mathbf{s}_t$ are assumed to be independent and identically distributed (i.i.d.) with unit power $\sigma^2_s$. Through multiplying $\mathbf{s}_t$ with the pre-equalization matrix $\mathbf{P}_t$ mentioned in \ref{subsec:framework}, the vector form of DD-domain ISAC symbols $\mathbf{x}_{t}$ can be obtained as 
\begin{equation}
    \mathbf{x}_t = \mathbf{P}_t \mathbf{s}_t.
\end{equation}

Based on the principle of OTFS modulation, the signal vector at the $t$-th time slot in time-delay (TD) domain $\mathbf{d}_{t}$ can be obtained by performing the inverse symplectic finite Fourier transform (ISFFT) and inverse discrete Fourier transform (IDFT) to ${{\mathbf{x}}_{t}}$. Mathematically, ${{\mathbf{d}}_{t}}$ can be expressed as
\begin{equation}
    {{\mathbf{d}}_{t}}=(\mathbf{F}_{N}^{H}\otimes {{\mathbf{I}}_{M}}){{\mathbf{x}}_{t}}.\label{eq1}
\end{equation}

The received sensing signal vector $\mathbf{r}_{S,t}$ at the AP in the TD domain is written as
\begin{align}
\label{eq:receive_sg}
\mathbf{r}_{S,t} =  {{{\mathbf{H}}}_{S,t}^{TD}} \mathbf{d}_{t} + \mathbf{w}_{S,t}, 
\end{align}
where $\mathbf{w}_{S,t}\in \mathbb{C}^{MN \times 1}$ is the additive white Gaussian noise (AWGN) vector at the AP, ${{\mathbf{H}}}_{S,t}^{TD}$ denotes the TD domain equivalent sensing channel in time slot $t$, respectively. According to \cite{9724198,thaj2022general}, the sensing channel is given by
\begin{align}
&{{\mathbf{H}}}_{S,t}^{TD}=\notag\\
&\sum_{p_S = 1}^{P_S} \!\!h_{t,p_S} e^{-j \frac{2 \pi{{l}}_{t,p_S} {{k}}_{t,p_S}}{M N}} \!\!\boldsymbol{\Delta}^{\left({{k}}_{t,p_S}\right)}\!\!\left(\!\boldsymbol{\Psi}^{\left({{l}}_{t,p_S}\right)}+\boldsymbol{\Psi}_{C P}^{\left({{l}}_{t,p_S}\right)}\!\right), \label{eq:H^TD_S}
\end{align}
where $P_S$ is the total number of sensing paths, ${{h}}_{t,p_S}$ is the channel attenuation corresponding to the $p_S$-th path. ${{l}}_{t,p_S}$ and ${{k}}_{t,p_S}$ are the delay tap and Doppler tap corresponding to the $p_S$-th path, respectively. Their relationship with the real delay ${\tau }_{t,p_S} = \frac{2R_{t,p_S}}{c}$ and Doppler shift ${\nu }_{t,p_S} = \frac{2f_0 v_{t,p_S}}{c}$ can be expressed as
    ${l}_{t,p_S}={{\tau }_{t,p_S}}{M\Delta f}, {k}_{t,p_S}={{\nu }_{t,p_S}}{NT}$, respectively,
where $R_{t,p_S}$ is the distance between the AP and the target, $c$ represents the speed of light, $f_0$ is the carrier frequency, $v_{t,p_S}$ is the radial velocity of the target relative to the AP, \(M\) is the number of delay grids, and \(N\) is the number of Doppler shift grids, $0\le {{l}_{t,p_S}}\le M-1$, $0\le {{\nu }_{t,p_S}}\le N-1$, $\Delta f$ is the subcarrier spacing, and $T$ is the time slot duration. In \eqref{eq:H^TD_S}, $\boldsymbol{\Delta}(p)$ characterizes as the Doppler influence, which can be expressed as$
\operatorname{diag}\left\{\eta^0, \eta^1, \ldots, \eta^{M N-1-l_{t,p}}, \eta^{-l_{t,p}} \ldots, \eta^{-1}\right\},
$
where $\eta \triangleq e^{\frac{j 2 \pi k_{t,p}}{M N}}$. 
\begin{align}
&\boldsymbol{\Psi}^{\left({{l}}_{t,p_S}\right)}=\notag\\ &\!\!\!\!\left(\begin{array}{@{\!}cccc@{\!}}
\psi\left(0-{{l}}_{t,p_S}\right) &\!\!\!\! \psi\left(-1-{{l}}_{t,p_S}\right) &\!\!\!\! \cdots &\!\!\!\! \psi\left(-b-{{l}}_{t,p_S}\right) \\
\psi\left(1-{{l}}_{t,p_S}\right) &\!\!\!\! \psi\left(0-{{l}}_{t,p_S}\right) &\!\!\!\! \ddots &\!\!\!\! \psi\left(-b+1-{{l}}_{t,p_S}\right) \\
\vdots &\!\!\!\! \ddots &\!\!\!\! \ddots &\!\!\!\! \vdots \\
\psi\left(b-{{l}}_{t,p_S}\right) &\!\!\!\! \psi\left(b-1-{{l}}_{t,p_S}\right) &\!\!\!\! \cdots &\!\!\!\! \psi\left(0-{{l}}_{t,p_S}\right)
\end{array}\right) \!\!,
\end{align}
where $b = MN-1$, and 
\begin{align}
    \boldsymbol{\Psi}_{C P}^{\left(l_i\right)}(i, j)= \begin{cases}\psi\left(b-(j-i)-l_i\right), & (j-i) \geq  b-L;\\ 0, & \text { else, }\end{cases}
\end{align}
where $L$ denotes the length of the cyclic prefix (CP). The $(i, j)$-th element of $\boldsymbol{\Psi}^{\left({{l}}_{t,p_S}\right)}$ is
\begin{align}
    \boldsymbol{\Psi}^{\left({{l}}_{t,p_S}\right)}(i, j)=\psi\left(i-j-{{l}}_{t,p_S}\right)=\operatorname{sinc}\left(i-j-{{l}}_{t,p_S}\right).
\end{align}

By performing Fourier transformation and symplectic finite Fourier transformation to the TD-domain received sensing signal $\mathbf{r}_{S,t}$, the DD-domain received sensing signal $\mathbf{y}_{S,t}$ at the $t$-th time slot is expressed as
\begin{align}
\mathbf{y}_{S,t}=\mathbf{H}_{S,t}^{DD}\mathbf{x}_{t}+\mathbf{z}_{S,t},
\end{align}
where $\mathbf{z}_{S,t}\in \mathbb{C}^{MN \times 1}$ is the DD-domain AWGN vector with power $\sigma^2_A$ at the AP, and $\mathbf{H}_{S,t}^{DD}$ is the DD-domain equivalent sensing channel in the $t$-th time slot, which can be expressed as
\begin{align}
\label{eq:DD_comm_chan}
    \mathbf{H}_{S,t}^{DD}=\left(\mathbf{F}_N \otimes \mathbf{I}_{M}\right){{\mathbf{H}}}_{S,t}^{TD}\left(\mathbf{F}_N^H \otimes \mathbf{I}_{M}\right).
\end{align}

\subsection{Communication Model}
The TD-domain received communication signal at UE is
\begin{align}
    \mathbf{r}_{C,t}={{{\mathbf{H}}}_{C,t}^{TD}}\mathbf{d}_t\!+\!\mathbf{w}_{C,t},
\end{align}
where $\mathbf{w}_{C,t}\in \mathbb{C}^{MN \times 1}$ is the i.i.d AWGN vector at the UE, ${{{\mathbf{H}}}_{C,t}^{TD}}$ denotes the communication channel in TD domain, expressed as
\begin{align}
\!\!&{{\mathbf{H}}}_{C,t}^{TD}=\notag\\
&\!\!\sum_{p_C = 1}^{P_C} \!\!h_{t,p_C} e^{-j \frac{2 \pi{{l}}_{t,p_C} {{k}}_{t,p_C}}{M N}} \!\!\boldsymbol{\Delta}^{\left({{k}}_{t,p_C}\right)}\!\!\left(\!\boldsymbol{\Psi}^{\left({{l}}_{t,p_C}\right)}\!\!+\!\!\boldsymbol{\Psi}_{C P}^{\left({{l}}_{t,p_C}\right)}\!\right),
\end{align}
where $P_C$ is the total number of communication paths. ${{h}}_{i,p_C}$, ${{l}}_{i,p_C}$, and ${{k}}_{i,p_C}$ are the channel attenuation, delay tap, and Doppler tap of the $p_C$-th path at the $t$-th time slot, respectively.  Similarly, by performing Fourier transformation and symplectic finite Fourier transformation to the TD-domain received signal, we have the DD-domain received communication signal $\mathbf{y}_{C,t}$ at the $t$-th time slot expressed as
\begin{align}
\mathbf{y}_{C,t}=\mathbf{H}_{C,t}^{DD}\mathbf{x}_{t}+\mathbf{z}_{C,t},
\end{align}
where ${{{\mathbf{H}}}_{C,t}^{DD}}$ is the communication channel in the DD domain, which can be expressed as
\begin{align}
    \mathbf{H}_{C,t}^{DD}=\left(\mathbf{F}_N \otimes \mathbf{I}_{M}\right){{\mathbf{H}}}_{C,t}^{TD}\left(\mathbf{F}_N^H \otimes \mathbf{I}_{M}\right),\label{eq:H_DD^C}
\end{align}
and $\mathbf{z}_{C,t}\in \mathbb{C}^{MN \times 1}$ is the i.i.d AWGN vector with power $\sigma^2_U$ in DD domain at UE.

\section{Problem Formulation}\label{sec:Problem}
In this section, we first analyze the performance metrics of interest, i.e., the CRLB of the sensing parameters (the real part and imaginary part of the channel attenuation, Doppler shift, and delay) and the MSE of the demodulated symbols. Based on the analysis, a multi-objective optimization problem is formulated to investigate the trade-off between sensing and communication performance via pre-equalization design.

\subsection{Performance Metrics}
\subsubsection{CRLB Analysis for Sensing Parameter Estimation}
CRLB is defined as the MSE lower bound of the unbiased estimate for the sensing parameters, i.e., 
\begin{align}
\mathrm{CRLB}_S \preceq \mathbb{E}[(\hat{\boldsymbol{\xi}}(\mathbf{y}_{S,t})-\boldsymbol{\xi})(\hat{\boldsymbol{\xi}}(\mathbf{y}_{S,t})-\left.\boldsymbol{\xi})^{\mathrm{H}}\right],
\end{align}
where $\boldsymbol{\xi}\in \mathbb{R}^{4 P_S}$ is the sensing parameter set to be estimated, and $\hat{\boldsymbol{\xi}}(\mathbf{y}_{S,t})$ is the estimated value of $\boldsymbol{\xi}$. In this work, the sensing parameters of interest include $\mathfrak{Re}\{h_{p_S}\}$, $\mathfrak{Im}\{h_{p_S}\}$, Doppler shift $\nu_{p_S}$, and delay $\tau_{p_S}$; hence, $\boldsymbol{\xi}(\mathbf{y}_{S,t})$ is expressed as
\begin{align}
    \boldsymbol{\xi}(\mathbf{y}_{S,t})=\left\lbrack \mathfrak{Re}\{h_{p_S}\},\mathfrak{Im}\{h_{p_S}\},\nu_{p_S}, \tau_{p_S} \right\rbrack_{1 \leq p_S \leq P_S}.
\end{align}

According to \cite{kay1993fundamentals}, the CRLB can be derived as the inverse of the Fisher Information Matrix (FIM) $\boldsymbol{\mathcal{I}}$, i.e.,
\begin{align}
   \mathrm{CRLB}_S = \boldsymbol{\mathcal{I}}^{-1}(\boldsymbol{\xi}(\mathbf{y}_{S,t})),
   \label{eq:invI}
\end{align}
where $\boldsymbol{\mathcal{I}}$ is expressed as
\begin{align}
\label{eq:FIM}
\boldsymbol{\mathcal{I}}(\boldsymbol{\xi}(\mathbf{y}_{S,t})) = \left\lbrack \mathcal{I}_{i,j} \right\rbrack_{1 \leq i,j \leq 4P_S}.
\end{align}
The element $\mathcal{I}_{i,j}$ of \eqref{eq:FIM} is
\begin{align}
\label{I_ij}
\mathcal{I}_{i,j} =2\mathfrak{R}\mathfrak{e}\left\{\!\! E\!\!\left[{\left( {\frac{\partial\mathbf{H}_{S,t}^{DD,{p_S}}}{\partial\xi_{j}}\mathbf{P}_t\mathbf{s}_t} \right)^{H}\mathbf{\Sigma}^{- 1}\frac{\partial\mathbf{H}_{S,t}^{DD,{p_S}}}{\partial\xi_{i}}\mathbf{P}_t\mathbf{s}_t} \right]\!\! \right\},
\end{align}
where $\xi_{i}$ and $\xi_{j}$ represent the $i$-th and $j$-th element of $\boldsymbol{\xi}(\mathbf{y}_{S,t})$, respectively, $\mathbf{\Sigma}= \frac{1}{\sigma^2_A} \mathbf{I}$ is the covariance matrix of $\mathbf{z}_{S,t}$, and \(\mathbf{H}_{S,t}^{DD,{p_S}}\) represents the channel matrix of the \(p_S\)-th sensing path. The derivation of the partial derivatives with respect to the relevant parameters $\frac{\partial\mathbf{H}_{S,t}^{DD,{p_S}}}{\partial\xi_{i}}$ and $\frac{\partial\mathbf{H}_{S,t}^{DD,{p_S}}}{\partial\xi_{j}}$ can be referred to in \cite{gaudio2020effectiveness} and is omitted here for brevity. For natation simplicity, \(\mathbf{H}_{S,t}^{DD,{p_S}}\) will be denoted as $\mathbf{H}^{{p_S}}$ by omitting the superscript $DD$ and subscripts $S$ and $t$, i.e.,
\begin{align}
    \mathbf{H}^{{p_S}} = \begin{bmatrix}
\mathbf{H}_{0,0}^{p_S} & \cdots & \mathbf{H}_{0,N - 1}^{p_S} \\
 \vdots & \ddots & \vdots \\
\mathbf{H}_{N - 1,0}^{p_S} & \cdots & \mathbf{H}_{N - 1,N - 1}^{p_S}
\end{bmatrix},
\end{align}
whose $(k',k)$-th element is denoted as
\begin{align}
    \mathbf{H}_{k^{'},k}^{p_S} = \begin{bmatrix}
{\mathbf{H}_{k^{'},k}^{p_S}\lbrack 0,0\rbrack} & \cdots & {\mathbf{H}_{k^{'},k}^{p_S}\lbrack 0,M - 1\rbrack} \\
 \vdots & \ddots & \vdots \\
{\mathbf{H}_{k^{'},k}^{p_S}\lbrack M - 1,0\rbrack} & \cdots & {\mathbf{H}_{k^{'},k}^{p_S}\lbrack M - 1,M - 1\rbrack}
\end{bmatrix}.
\end{align}

To further simplify the expression of the CRLB, we rewrite $\mathcal{I}_{i,j}$ in \eqref{I_ij} as
\begin{align}
 \mathcal{I}_{i, j}& \mathop=\limits^{(a)} \frac{2}{\sigma^2_A}\mathfrak{R e}\left\{E\left[\left(\frac{\partial \mathbf{H}^{{p_S}}}{\partial \xi_j} \mathbf{P}_t \mathbf{s}_t\right)^{\mathrm{H}} \frac{\partial \mathbf{H}^{{p_S}}}{\partial \xi_i} \mathbf{P}_t \mathbf{s}_t\right]\right\} \notag\\
 &=\frac{2}{\sigma^2_A} \mathfrak{R e}\left\{E\left[\operatorname{tr}\left(\left(\frac{\partial \mathbf{H}^{{p_S}}}{\partial \xi_j} \mathbf{P}_t \mathbf{s}_t\right)^{\mathrm{H}} \frac{\partial \mathbf{H}^{{p_S}}}{\partial \xi_i} \mathbf{P}_t \mathbf{s}_t\right)\right]\right\} \notag \\ 
  &=\frac{2}{\sigma^2_A} \mathfrak{R e}\left\{E\left[\operatorname{tr}\left(\mathbf{s}_t^H\mathbf{P}_t^H\left( \frac{\partial \mathbf{H}^{{p_S}}}{\partial \xi_j}\right)^H   \frac{\partial \mathbf{H}^{{p_S}}}{\partial \xi_i} \mathbf{P}_t \mathbf{s}_t\right)\right]\right\} \notag \\ 
 & \mathop=\limits^{(b)}\frac{2}{\sigma^2_A} \mathfrak{R e}\left\{\operatorname{tr}\left(\frac{\partial \mathbf{H}^{{p_S}}}{\partial \xi_i} \mathbf{P}_t E\left[\mathbf{s}_t\mathbf{s}_t^H\right] \mathbf{P}_t^H\left(\frac{\partial \mathbf{H}^{{p_S}}}{\partial \xi_j}\right)^H\right)\right\} \notag \\ 
 & \mathop=\limits^{(c)}\frac{2 \sigma^2_s}{\sigma^2_A} \mathfrak{R e}\left\{\operatorname{tr}\left(\frac{\partial \mathbf{H}^{{p_S}}}{\partial \xi_i} \mathbf{P}_t \mathbf{P}_t^H\left(\frac{\partial \mathbf{H}^{{p_S}}}{\partial \xi_j}\right)^H\right)\right\},
 \label{eq:I_ij}
\end{align}
where the step (a) leverages the i.i.d. property of AWGN, with its covariance matrix represented as $\sigma^2_A \mathbf{I}$. The derivation of step (b) makes use of the circular nature of matrix trace. The derivation of step (c) is based on the assumption that the modulated symbols in $\mathbf{s}_t$ are i.i.d. with unit power $\sigma^2_s$.

Combing \eqref{eq:I_ij} with \eqref{eq:FIM} and then substituting back to \eqref{eq:invI}, we can obtain the CRLB of $\boldsymbol{\xi}(\mathbf{y}_{S,t})$.

\subsubsection{MSE Analysis of Demodulated Symbols}
Most of the previous works on ISAC waveform optimization adopted SINR-based metrics for evaluating communication performance, such as the sum rate. However, these metrics are unsuitable for the proposed pre-equalization framework. This is because, although such metrics can achieve optimal reception from the power perspective, they do not account for the constellation angle distortion under the direct symbol demodulation scheme employed at the UE. Therefore, in this paper, a more suitable metric, namely, MSE of communication symbols, is adopted to evaluate communication performance.

MSE of communication symbols is defined as the mean squared error between the DD-domain received communication signal $\mathbf{y}_{C,t}$ and the modulated symbol vector $\mathbf{s}_t$. Mathematically, it is expressed by
\begin{align}
    \mathrm{MSE}_C = \mathbb{E}\left[(\beta\mathbf{y}_{C,t}-\mathbf{s}_t)^{\mathrm{H}}(\beta\mathbf{y}_{C,t}-\mathbf{s}_t)\right],
\end{align}
where $\beta$ is the power normalization factor used to normalize the power of the received communication signal $\mathbf{y}_{C,t}$ to the power of the modulated symbol vector $\mathbf{s}_t$.

\subsection{Optimization Problem Formulation}
To achieve a trade-off between communication and sensing performance, we consider using the weighted sum of the two metrics as the optimization objective. By properly setting the weights of communication and sensing performance, pre-equalization design can be tailored to meet various communication and sensing requirements. Specifically, the pre-equalization design problem can be formulated as
\begin{align}
\label{optproblem}
(\mathcal{P}): & \min _{\mathbf{P}_t} \rho_{C} \mathrm{MSE}_C + (1-\rho_{C}) \mathrm{CRLB}_S, \\
&\label{optproblemcons} \text { s.t. } \|\mathbf{P}_t\|_F^2 \leq P_{\max},
\end{align}
where $\rho_{C}$ is the weighting factor of communication performance, $(1-\rho_{C})$ is the weighting factor of sensing communication, and $P_{\mathrm{max}}$ is the maximum transmission power, which is limited by the hardware of the AP.

The complicated objective makes the formulated problem \eqref{optproblem} challenging to solve. First, since the optimization variable is continuous, the optimal solution cannot be obtained via exhaustive searching. In addition, this work considers high-mobility scenarios where instantaneous perfect CSI and sensing parameters are unavailable, leading to the following problems: On one hand, the pre-equalization design based on the outdated CSI results in suboptimal results.  On the other hand, the computation of the CRLB relies on true sensing parameters, which are unavailable before sensing parameter estimation. Note that such a critical issue has been largely overlooked in existing optimization works.

\vspace{-0mm}\section{Proposed Algorithms}\label{sec:algorithm}
In this section, we present our developed algorithm for solving the optimization problem in \eqref{optproblem}. We first describe the overall CP-PE framework, which is then followed by a detailed description of the channel prediction module, CSI processing module, and pre-equalization module.
\subsection{Overall Framework}
Recently, deep learning has demonstrated significant capabilities in solving complicated problems. It can leverage the offline training mechanism to improve the generalization ability of the network; in the online deployment phase, only one low-latency forward propagation is required to obtain the pre-equalization matrix solution, effectively addressing rapidly changing channel conditions. Therefore, the deep learning approach is adopted in this work to solve the original optimization problem.

\begin{figure*}
    \centering
    \includegraphics[width=0.9\linewidth]{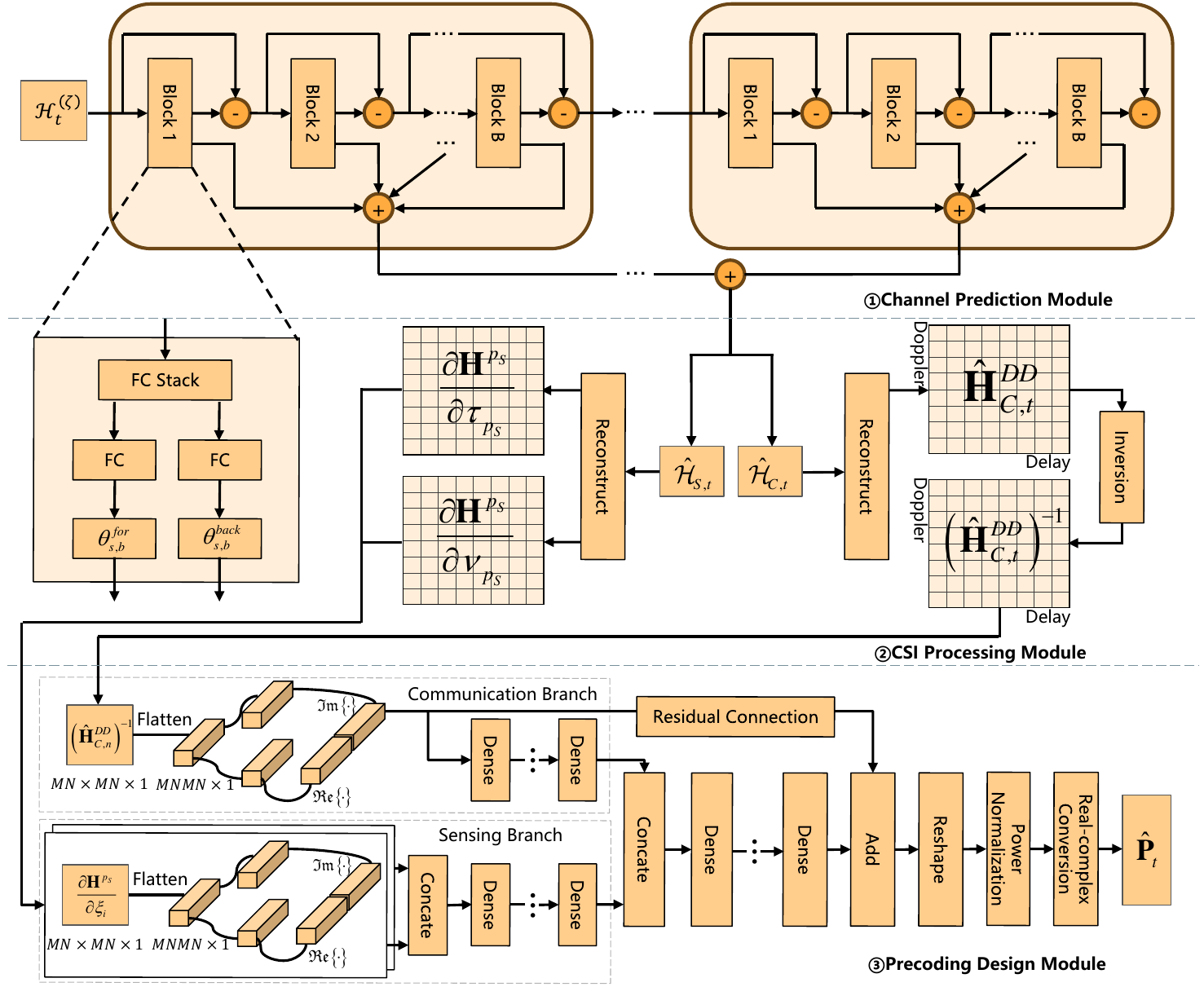}
    \caption{Overall framework of the channel prediction based pre-equalization design.}
    \label{fig:algorithm}
\end{figure*}

The overall framework is illustrated in Fig. \ref{fig:algorithm}. Specifically, to address the outdated CSI and sensing parameters in high-mobility scenarios, we first develop a communication and sensing channel parameter prediction network to forecast the channel coefficients in the delay-Doppler domain. By leveraging the temporal dependence of OTFS channel parameters, the channel parameters in the next time slot can be predicted. This not only ensures that the pre-equalization design is based on almost real-time CSI and sensing parameters but also reduces the overhead for obtaining CSI. Based on the predicted CSI and sensing parameters, the AP facilitates pre-equalization design with DNN, which enables direct demodulation of signals at the UAV. 

Note that instead of merging channel parameter prediction with pre-equalization, we decompose the above process into three modules, i.e., channel prediction module, CSI reconstruction module, and pre-equalization design module. This comes from the fact that the three modules of the network can be trained independently, thereby reducing the overall training complexity of the network. In the channel prediction module, based on the sparsity of the OTFS channel and the strong correlation between the OTFS channel and mobility, only a few channel parameters, such as channel attenuation values, delays, and Doppler shifts, are required. After inputting the above parameters in the past few slots into the prediction network, the channel and sensing parameters for the next time slot can be predicted. These predicted parameters are then fed into the CSI processing module to be processed and then provide an initial value for the pre-equalization design module, thereby reducing the complexity of the pre-equalization network training process. 

\vspace{-0mm}\subsection{Channel Prediction Module}
To design the pre-equalization matrix in problem \eqref{optproblem}, we first need to obtain the CSI and the sensing parameters of the $t$-th time slot. To effectively reduce the overhead of channel estimation and reduce the impact of outdated CSI, we come up with a channel prediction module based on the neural basis expansion analysis for interpretable time series forecasting (NBEATS) network \cite{oreshkin2019n}, which relies on the channel information of the past few time slots. 

The core idea of the module is as follows. First, the prediction module leverages the sparsity of OTFS channels to decouple the channel into a series of parameters. This enables a parameter-level channel prediction using a low-dimensional input space, thereby reducing complexity. Furthermore, it decomposes time series through multiple fully connected layers, with each layer fitting partial features of the channel sequence. This approach allows the network to extract, separate, and integrate channel variations from multiple scales. Consequently, the prediction does not rely on period characteristics, which are difficult to maintain for channel parameters strongly associated with mobility. This enables the effective fitting of rapidly changing channel features with fewer time steps, even in the absence of clear periodicity. The details of the module are described below.

\subsubsection{Data Pre-processing}
Leveraging the sparsity of OTFS channels, the channel parameters, i.e., delays ${{\tau }_{t}}$, Doppler shifts ${{\nu }_{t}}$ and channel attenuation parameters ${{h}}_{t}$, are adopted as the input of the prediction network instead of the whole channel matrix as considered in \cite{liu2023predictive}. This greatly reduces the dimensionality of the network input, effectively lowering the network overhead. In this way, the communication channel parameter set at the $t$-th time slot is denoted as
\begin{align}
    \mathcal{H}_{C,t} = \left\{\mathfrak{Re}\left\{{h}_{t} \right\}, \mathfrak{Im}\left\{{h}_{t} \right\}, {{\tau }_{t}}, {{\nu }_{t}}\right\}.
\end{align}
Similarly, the sensing channel parameter set can be represented as $\mathcal{H}_{S,t}$. For brevity, the subscripts denoting communication and sensing in $\mathcal{H}_{C,t}$ and $\mathcal{H}_{S,t}$ are omitted in the following.

To predict the channel parameter set at the $t$-th time slot, the channel parameter sets $\mathcal{H}^{(\zeta)}_{t}$ of the past $\zeta$ time slots are fed into the prediction network, where $\mathcal{H}^{(\zeta)}_{t}$ is expressed as
\begin{align}
    \mathcal{H}^{(\zeta)}_{t} = \left\{\mathcal{H}_{t-\zeta},\mathcal{H}_{t-\zeta+1},...,\mathcal{H}_{t-1}\right\}.
\end{align}

\subsubsection{Network Structure}
The adopted prediction network structure is illustrated in Fig. \ref{fig:algorithm} \textcircled{1}.

The channel prediction network consists of several stacks, which are connected as shown in Fig. \ref{fig:algorithm} \textcircled{1}. Each stack extracts one type of feature from the channel parameter variations and passes the unlearned feature variations to the next stack for processing, ultimately achieving the learning and prediction of channel parameter features at different scales. The channel parameter prediction results of each stack are summed up to form the predicted channel parameter set $\mathcal{\hat{H}}_{t}$, i.e., 
\begin{align}
    \mathcal{\hat{H}}_{t} = \left\{\mathfrak{Re}\left\{{\hat{h}}_{t} \right\}, \mathfrak{Im}\left\{{\hat{h}}_{t} \right\}, {{\hat{\tau} }_{t}}, {{\hat{\nu} }_{t}}\right\}.
\end{align}

Each stack is composed of $B$ blocks. For each block, the input is a channel parameter sequence $\mathcal{H}^{(\zeta)}_{t,\ell}$ of length \(\zeta\), containing a portion of the channel features. The output consists of two parts, i.e., one part is a fitted channel parameter sequence of length $\zeta$, representing the features of the channel parameter sequence that the block can fit, and the other part is the predicted channel parameter value $\widehat{\mathcal{H}}_{t}$ for the next slot. Note for the first block in the entire model, its input is the original channel parameter sequence $\mathcal{H}^{(\zeta)}_{t}$. 

Different blocks are connected based on the doubly residual stacking principle \cite{oreshkin2019n}. In other words, each block $\ell$ contains two residual branches, i.e., one branch computes the difference between the fitted channel parameter sequence and the block's input, outputting the channel parameter sequence that the block failed to fit, which is then passed to the next block for fitting. The other branch sums the predicted channel parameter value $\widehat{\mathcal{H}}_{t,\ell}$ of the block $\ell$ with those from other blocks, serving as the channel parameter prediction output for the stack. This process is represented as
\begin{align}
    \mathcal{H}^{(\zeta)}_{t,\ell} = \mathcal{H}^{(\zeta)}_{t,\ell-1} - \widehat{\mathcal{H}}^{(\zeta)}_{t,\ell-1},
\end{align}
\begin{align}
    \widehat{\mathcal{H}}_{t} = \sum_{\ell} \widehat{\mathcal{H}}_{t,\ell}.
\end{align}

Inside each block $\ell$, the network is composed of two parts. The first part is a 4-layer fully connected network with a channel parameter sequence $\mathcal{H}^{(\zeta)}_{t,\ell}$ as input. The output of this fully connected network is input into two other fully connected networks. One of these fully connected networks is responsible for fitting the input channel parameter sequence $\widehat{\mathcal{H}}^{(\zeta)}_{t,\ell}$ of length $\zeta$. The output result is linearly mapped to output the fitted channel parameter sequence $\widehat{\mathcal{H}}^{(\zeta)}_{t,\ell}$. The other fully connected network's output is linearly mapped to predict the channel parameters $\widehat{\mathcal{H}}_{t,\ell}$ for the next time slot. The linear mapping process is expressed as follows:
\begin{align}
    \widehat{\mathcal{H}}_{t,\ell} = \mathbf{V}_{\ell}^f \theta_{\ell}^f + \mathbf{b}_{\ell}^f, 
\end{align}
\begin{align}
    \widehat{\mathcal{H}}^{(\zeta)}_{t,\ell} = \mathbf{V}_{\ell}^b \theta_{\ell}^b + \mathbf{b}_{\ell}^b,
\end{align}
where $\mathbf{V}_{\ell}^b$ and $\mathbf{V}_{\ell}^f$ are the learnable weight matrices for channel parameter fitting and prediction. $\theta_{\ell}^b$ and $\theta_{\ell}^f$ represent the output of the fully connected layers for channel parameter fitting and prediction.  $\mathbf{b}_{\ell}^b$ and $\mathbf{b}_{\ell}^f$ are the learnable bias vectors for channel parameter fitting and prediction, respectively.

\subsubsection{Implementation Process}
For the channel prediction module, the network implementation process is divided into two stages, i.e., offline training and online prediction. In the offline training stage, a large set of channel time series data is used to train the network. Notably, the channel parameter dataset is generated using ray-tracing based on the UE's trajectory, with periodic sampling over a continuous period. In this way, more realistic channel parameters are collected, thereby ensuring the practical applicability of the developed network.

The loss function of the offline training process is designed based on the MSE between predicted channel parameters and real channel parameters, denoted as
\begin{align}
    \mathcal{L}1= \| \mathcal{H}_{t} - \hat{\mathcal{H}}_{t} \|_2^2.
\end{align}
Subsequently, we consider using six stacks, each containing three blocks in our channel prediction module. We utilized the Adam optimizer to iteratively train the channel prediction network with a learning rate of 0.001 until convergence. 

During the online prediction phase, simply input the channel parameters $\mathcal{H}^{(\zeta)}_{t}$ estimated from the previous time step into the channel prediction module. Then, a single forward computation is done to predict the channel parameters $ \mathcal{\hat{H}}_{t}$ for the current time slot $t$.

\subsection{CSI Processing Module}
The CSI processing module is mainly designed to post-process the output of the channel prediction module and provide an initial value for the pre-equalization design module, thereby reducing the complexity of learning and training. First, the predicted channel parameters from the channel prediction module, including channel attenuation, delay, and Doppler shift, are used to reconstruct the communication channel $\mathbf{\hat{H}}_{C,t}^{DD}$ according to \eqref{eq:H_DD^C}. Subsequently, the inverse of the communication channel matrix  \(\left(\mathbf{\hat{H}}_{C,t}^{DD}\right)^{-1}\) is computed.

\begin{table*}[!t]
\centering
\caption{Multi-objective DNN network structure}
\label{tab:DNN}
\begin{tabular}{|c|c|c|c|}
\hline
\textbf{Layer} & \textbf{Activation} & \textbf{Output shape} & \textbf{Connected to} \\ \hline

\multicolumn{4}{|c|}{\textbf{Input and Pre-processing}} \\ \hline
Input layer 1 ($\mathbf{H}_{C,t}^{-1}$) &  & $(2M\times N, M\times N)$ & $[\mathfrak{Re}\{\mathbf{(\hat{H}}_{C,t}^{DD})^{-1}\}, \mathfrak{Im}\{\mathbf{(\hat{H}}_{C,t}^{DD})^{-1}\}]$ \\ \hline
Input layer 2 ($\dfrac{\partial \mathbf{H}^{p_S}}{\partial \tau_{p_S}}$) &  & $(2M\times N, M\times N)$ & $[\mathfrak{Re},\mathfrak{Im}]\{\dfrac{\partial \mathbf{H}^{p_S}}{\partial \tau_{p_S}}\}$ \\ \hline
Input layer 3 ($\dfrac{\partial \mathbf{H}^{p_S}}{\partial \nu_{p_S}}$) &  & $(2M\times N, M\times N)$ & $[\mathfrak{Re},\mathfrak{Im}]\{\dfrac{\partial \mathbf{H}^{p_S}}{\partial \nu_{p_S}}\}$ \\ \hline
Flatten layers &  & $(2M\times N\times M\times N)$ & Each input layer \\ \hline

\multicolumn{4}{|c|}{\textbf{Communication Branch}} \\ \hline
Dense layer 1 (comm\_dense1) & ReLU & $128$ & Flatten($\mathbf{H}_{C,t}^{-1}$) \\ \hline
Dropout (comm\_drop1) &  & $128$ & Dense layer 1 \\ \hline
Dense layer 2 (comm\_dense2) & ReLU & $256$ & Dropout (comm\_drop1) \\ \hline
Dropout (comm\_drop2) &  & $256$ & Dense layer 2 \\ \hline
Dense layer 3 (comm\_dense3) & ReLU & $512$ & Dropout (comm\_drop2) \\ \hline
Dense layer 4 (comm\_output) & Linear & $M\times N\times M\times N$ & Dense layer 3 \\ \hline

\multicolumn{4}{|c|}{\textbf{Sensing Branch}} \\ \hline
Concatenate (sensing\_concat) &  & $2\times(2M\times N\times M\times N)$ & Flatten($\dfrac{\partial \mathbf{H}^{p_S}}{\partial \tau_{p_S}}$) + Flatten($\dfrac{\partial \mathbf{H}^{p_S}}{\partial \nu_{p_S}}$) \\ \hline
Dense layer 1 (sensing\_dense1) & ReLU & $128$ & Concatenate layer \\ \hline
Dropout (sensing\_drop1) &  & $128$ & Dense layer 1 \\ \hline
Dense layer 2 (sensing\_dense2) & ReLU & $256$ & Dropout (sensing\_drop1) \\ \hline
Dropout (sensing\_drop2) &  & $256$ & Dense layer 2 \\ \hline
Dense layer 3 (sensing\_dense3) & ReLU & $512$ & Dropout (sensing\_drop2) \\ \hline
Dense layer 4 (sensing\_output) & Linear & $M\times N\times M\times N$ & Dense layer 3 \\ \hline

\multicolumn{4}{|c|}{\textbf{Fusion and Output Module}} \\ \hline
Concatenate (fusion\_layer) &  & $2M\times N\times M\times N$ & comm\_output + sensing\_output \\ \hline
Dense layer 1 (fusion\_dense1) & ReLU & $128$ & Fusion layer \\ \hline
Dropout (fusion\_drop1) &  & $128$ & Dense layer 1 \\ \hline
Dense layer 2 (fusion\_dense2) & ReLU & $256$ & Dropout (fusion\_drop1) \\ \hline
Dropout (fusion\_drop2) &  & $256$ & Dense layer 2 \\ \hline
Dense layer 3 (fusion\_dense3) & ReLU & $512$ & Dropout (fusion\_drop2) \\ \hline
Dense layer 4 (final\_output) & Linear & $2M\times N\times M\times N$ & Dense layer 3 \\ \hline
Add layer (residual) &  & $2M\times N\times M\times N$ & Flatten($\mathbf{H}_{C,t}^{-1}$) + final\_output \\ \hline
Reshape layer &  & $(2M\times N, M\times N)$ & Add layer \\ \hline
\end{tabular}
\end{table*}

\vspace{-0mm}\subsection{Pre-equalization Design Module}
The output of the CSI processing module and sensing parameters can then be used to aid the pre-equalization matrix design. Therein, we leverage a neural network to design the pre-equalization matrix as shown in Fig. \ref{fig:algorithm} \textcircled{3}. The details of the module are described below.

\subsubsection{Data Pre-processing}
The input of the proposed multi-objective optimization network consists of the inverse matrix of the predicted communication channel and the sensing derivative matrices $\dfrac{\partial \mathbf{H}^{p_S}}{\partial \tau_{p_S}}$ and $\dfrac{\partial \mathbf{H}^{p_S}}{\partial \nu_{p_S}}$. 
For each input, the real and imaginary parts are separated and concatenated along dimension~1 to ensure compatibility with real-valued neural network operations. 
Specifically, the real and imaginary parts of the inverse communication channel matrix are combined to form a tensor of size $(2M \times N, M \times N)$, and each of the derivative matrices $\dfrac{\partial \mathbf{H}^{p_S}}{\partial \tau_{p_S}}$ and $\dfrac{\partial \mathbf{H}^{p_S}}{\partial \nu_{p_S}}$ is processed in the same manner. 
Each tensor is then flattened into one-dimensional vectors, denoted as $\mathbf{t}_{H}$, $\mathbf{t}_{\tau_{p_S}}$, and $\mathbf{t}_{\nu_{p_S}}$. 
These vectors are fed into different branches of the network corresponding to the communication and sensing objectives.

\subsubsection{Network Structure}
The proposed pre-equalization design network adopts a dual-branch architecture to jointly optimize communication and sensing performance. 
The communication branch takes the flattened inverse matrix of the predicted communication channel as input and processes it through several dense layers with ReLU activation and dropout for regularization. 
The sensing branch receives the concatenated flattened vectors of $\dfrac{\partial \mathbf{H}^{p_S}}{\partial \tau_{p_S}}$ and $\dfrac{\partial \mathbf{H}^{p_S}}{\partial \nu_{p_S}}$, which are then processed through a stack of dense layers. 
Through this dual-branch architecture, features closely related to pre-equalization are extracted separately from communication and sensing parameters, thereby reducing the processing complexity of subsequent pre-equalization matrix optimization.

Outputs of the two branches are concatenated in a fusion layer, which aggregates features from both domains. The fused features are further processed by dense layers to generate the pre-equalization output. To accelerate convergence and retain model-based priors, a residual connection is introduced between the flattened inverse channel matrix and the output. This enables the network to refine the model-based prior by directly adding learned corrections. Finally, the output is reshaped into a matrix of size \((2M \times N, M \times N)\), corresponding to the real and imaginary parts of the pre-equalization matrix before power normalization. The detailed parameters of the proposed DNN-based pre-equalization design network are listed in TABLE \ref{tab:DNN}.

\subsubsection{Data Post-processing}
The output of the last reshape layer is a matrix with a size of \((2M\times N, M\times N)\), which is denoted as \(\bar{\mathbf{P}}_{t}\). The first \(MN\) rows \(\bar{\mathbf{P}}_{t}(1:MN,:)\) correspond to the real part of the pre-equalization matrix before power normalization, and the subsequent \(MN\) rows \(\bar{\mathbf{P}}_{t}(MN+1:2MN,:)\) correspond to the imaginary part.

Due to the transmission power constraint in \eqref{optproblemcons}, power normalization is applied as
\begin{align}
    {\tilde{\mathbf{P}}}_{t}=\sqrt{P_{\text{max}}} \frac{\bar{\mathbf{P}}_{t}}{\left\|\bar{\mathbf{P}}_{t}\right\|_F}.
\end{align}

Then, the real-value outputs are reconnected into the complex-valued pre-equalization matrix as
\begin{align}
\hat{\mathbf{P}}_{t} = \left[\tilde{\mathbf{P}}_{t}(1:MN,:) + j\,\tilde{\mathbf{P}}_{t}(MN+1:2MN,:)\right].
\end{align}

\subsubsection{Implementation Process}
Similar to the channel prediction module, the implementation process involves two phases, i.e., offline training and online deployment. During the offline training phase, to minimize the objective problem in \eqref{optproblem}, we design the following loss function
\begin{align}
    \mathcal{L}2= \rho_{C} MSE_C + (1-\rho_{C}) \mathrm{CRLB}_S + \rho_{L} \Phi(\mathbf{\psi}_2),
\end{align}
where $\Phi(\mathbf{\psi}_2)$ is the L2 regularization term, which is used to prevent the model from overfitting. $\mathbf{\psi}_2$ is the parameter of the DNN-based pre-equalization design network. $\rho_{L}$ is the weighting factor of the L2 regularization term. 

The channel information processed by the CSI processing module is used as input for the network, and the sensing parameters are used as labels in the loss function. Based on them, the neural network is then trained offline using the Adamax optimizer with the loss function $\mathcal{L}2$ and a learning rate of 0.001 until the network converges. 

Due to the well-established offline training, during the online deployment phase, only one forward propagation is required to complete the prediction of the pre-equalization matrix for the next time slot considering both communication and sensing performance. Therefore, the pre-equalization matrix can be rapidly obtained by inputting the inverse of the channel matrix and the modulated symbols. After using $\hat{\mathbf{P}}_{t}$ to pre-equalize the transmitted symbols, the sensing performance of the transmitted waveform can be improved, and at the same time, the UE can also directly demodulate the signal without equalization, thus significantly reducing the complexity of the UE.

\vspace{-0mm}\section{Simulation Results}\label{sec:result}
 In this section, the numerical results are presented to demonstrate the effectiveness of the proposed scheme. We consider one AP, one UE, which also serves as a sensing target, and $C=2$ scatters. The AP is fixed at coordinate $(0 \mathrm{m},0 \mathrm{m})$, while the UE moves around the AP at speeds of up to $95 \mathrm{m/s}$. These positions and motion patterns are incorporated into the Wireless Insite \cite{remcom_wireless_insite} software to generate corresponding channels as the dataset, including delays, Doppler shifts, and channel attenuation parameters. The dataset consists of 640 trajectory groups, which are divided into training and testing sets in a ratio of 4:1. Moreover, the size of one OTFS symbol is set as $M=16$ and $N=16$\cite{10915679, 10192916, 11164940, 8892482}, respectively. The length of the CP is $L=4$. The speed of light is \( c = 3 \times 10^{8}~\mathrm{m/s} \). 
The carrier frequency is \( f_0 = 3~\mathrm{GHz} \), and the subcarrier spacing is \( \Delta f = 15~\mathrm{kHz} \). 
Accordingly, the time slot duration is given by 
$T = 66.67~\mu\mathrm{s}$.

To verify the performance of the proposed CP-PE algorithm, we compare our algorithm with the following schemes, i.e., 

\begin{itemize}
\item {\textit{Minimum mean square error (MMSE) equalization\cite{hong2022delay}: }} Under this scheme, the MMSE equalization, which is calculated using the communication channel matrix of perfect CSI, is performed at the UE side. In this work, MMSE is used as an upper-bound algorithm for communication performance. 
\item {\textit{Pre-equalization with perfect CSI:}} This is the ideal case, where only the pre-equalization design network proposed by this paper is utilized to design the pre-equalization matrix. The network takes the communication channel matrix with perfect CSI as input.
\item {\textit{Pre-equalization with outdated CSI:}} Under this scheme, channel prediction is not performed, and the CSI from the previous time slot is used as input to the pre-equalization network.
\end{itemize}

\subsection{Performance of Channel Prediction}
\begin{figure}[!t]
\centering
\includegraphics[width=\linewidth]{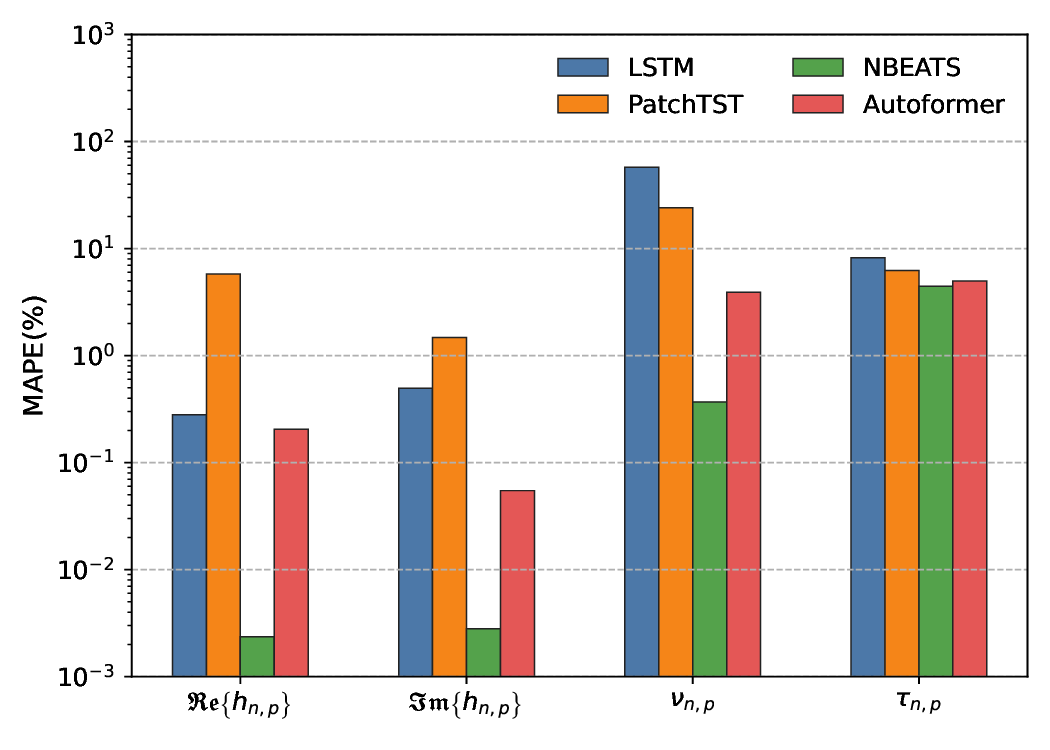}
\caption{OTFS channel parameter prediction accuracy comparison.}
\label{fig:predict_MAPE}
\end{figure}

We first examine the performance of channel prediction. Fig. \ref{fig:predict_MAPE} compares the accuracy of the NBEATS-based OTFS channel prediction module with the common time-series prediction algorithm, including long short-term memory recurrent neural network (LSTM) \cite{yu2019review}, patch time series Transformer (PatchTST) \cite{nie2022time}, and Autoformer \cite{wu2021autoformer}. Due to the varying magnitudes of OTFS channel parameters, the mean absolute percentage error (MAPE) is employed as the accuracy evaluation metric, expressed as
\begin{align}
\mathrm{MAPE} = \frac{1}{S} \sum_{i=1}^{S} \left| \frac{z_i - \hat{z}_i}{z_i} \right| \times 100,
\end{align}
where \(z_i\) and \(\hat{z}_i\) are the true and predicted values of the parameters, respectively, and \(S\) is the number of samples.

As shown in the figure, for the prediction of \(\mathfrak{Re}\left(h_{t,p_C}\right)\), \(\mathfrak{Im}\left(h_{t,p_C}\right)\), \(\nu_{t,p_C}\), and \(\tau_{t,p_C}\), the NBEATS algorithm always demonstrates the best MAPE performance. The worst prediction performance for the NBESTS algorithm is observed for the delay \(\tau_{t,p_C}\), while its MAPE is still below 5\%. In addition, for the predictions of \(\mathfrak{Re}\left(h_{t,p_C}\right)\) and \(\mathfrak{Im}\left(h_{t,p_C}\right)\), the MAPE is significantly less than 1\%. Overall, this demonstrates that selecting the NBEATS algorithm for OTFS channel parameter prediction is feasible. 

\subsection{Convergence Behavior}
\begin{figure}[t]
\centering
\subfigure[Communication MSE performance.]{\label{fig:MSE_C5S0_01_diff_iter}\includegraphics[width=\linewidth]{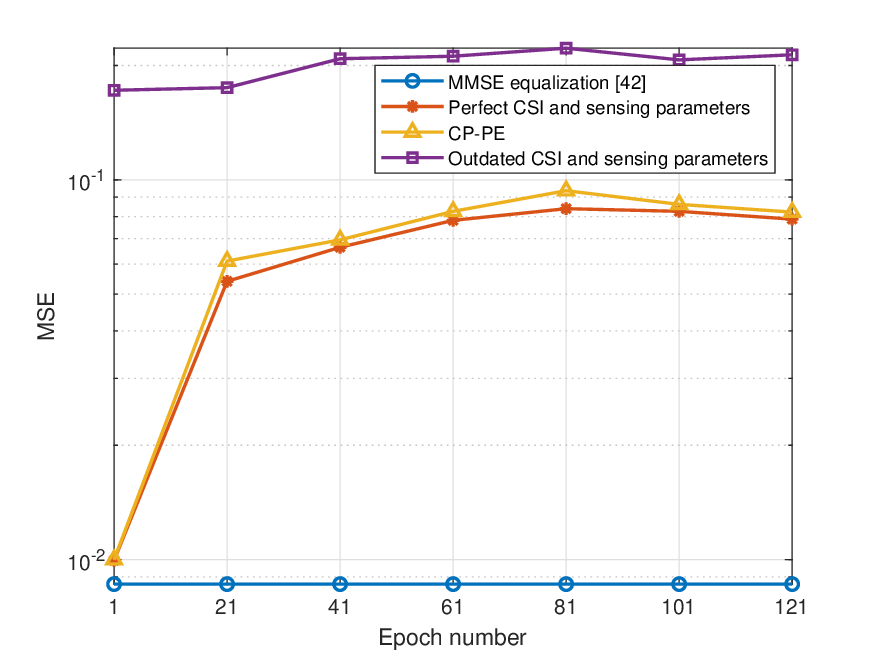}}\\
\subfigure[Sensing CRLB performance.]{\label{fig:CRB_C5S0_01_diff_iter}\includegraphics[width=\linewidth]{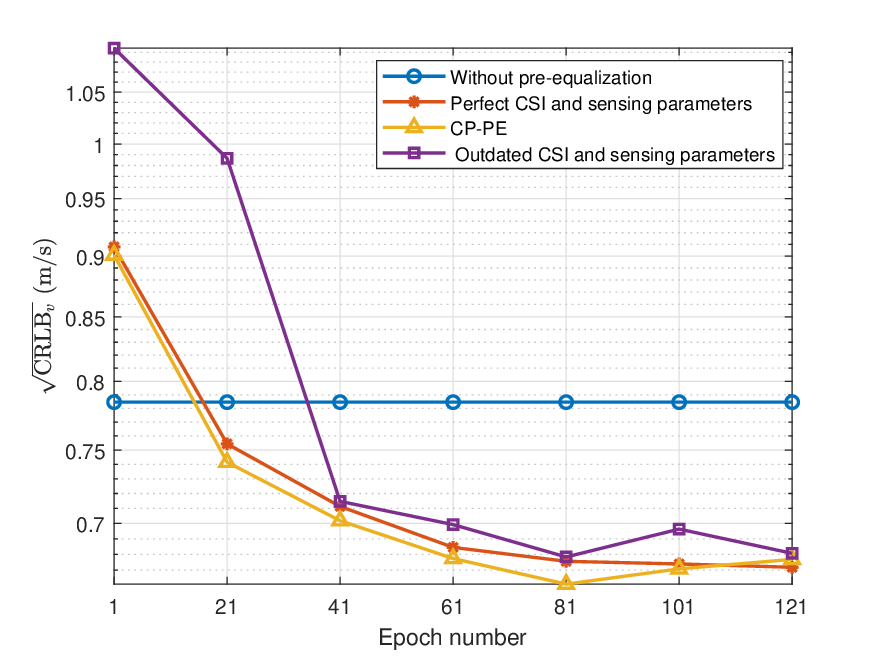}}
\caption{{Sensing CRLB and communication MSE performance under different iteration numbers.}}\label{fig:diff_iter}
\end{figure}

This subsection takes the case where communication and sensing performances are equally weighted (i.e., $\rho_C = 0.5$, $\rho_S = 0.5$) as an example and explores the relationship between the number of iterations during the offline training phase of the pre-equalization design network and the ISAC performance. For communication performance, as shown in Fig. \ref{fig:MSE_C5S0_01_diff_iter}, the MSE gradually converges from a level close to that of the MMSE equalizer to around 0.08. This is because MMSE equalization at the receiver represents the optimal communication performance, while considering sensing performance inevitably leads to a certain degradation in communication MSE. For sensing performance, as shown in Fig. \ref{fig:CRB_C5S0_01_diff_iter}, $\sqrt{\mathrm{CRLB}_v}$ decreases progressively from a value higher than that without pre-equalization to below 0.7 m/s, demonstrating the effectiveness of the proposed pre-equalization design in improving sensing accuracy. Notably, CP-PE can occasionally outperform the case with perfect CSI and sensing parameters, as it optimizes the weighted joint communication–sensing performance. Due to the inherent trade-off between communication and sensing, performance crossover between two algorithms may arise when CRLB is evaluated individually. In both figures, the performance obtained using the outdated pre-equalization matrix is inferior to that based on channel prediction, particularly in terms of communication MSE. This observation indicates that accurate channel information is crucial for the proper operation of the pre-equalization framework.

From the convergence behavior, it can be observed that the training tends to converge after approximately 60 epochs. This convergence mainly corresponds to the optimization of sensing performance, as the communication performance is already close to optimal at the early stage of training. Therefore, when communication performance is prioritized in the OTFS-based ISAC system, a large number of training iterations may not be necessary to meet the performance requirements. However, if a strict trade-off between communication and sensing performance is desired, additional iterations are required to achieve convergence in both objectives. Although the offline training phase takes some iterations, during online deployment, the designed CP-PE network does not require as many iterations and only a single forward propagation is required to complete the design of the pre-equalization matrix considering both communication and sensing performance. The ability of this network to provide rapid outputs helps overcome outdated pre-equalization matrix designs in high-mobility scenarios.

\subsection{Impact of Transmit Power}
\begin{figure}[t]
\centering
\includegraphics[width=\linewidth]{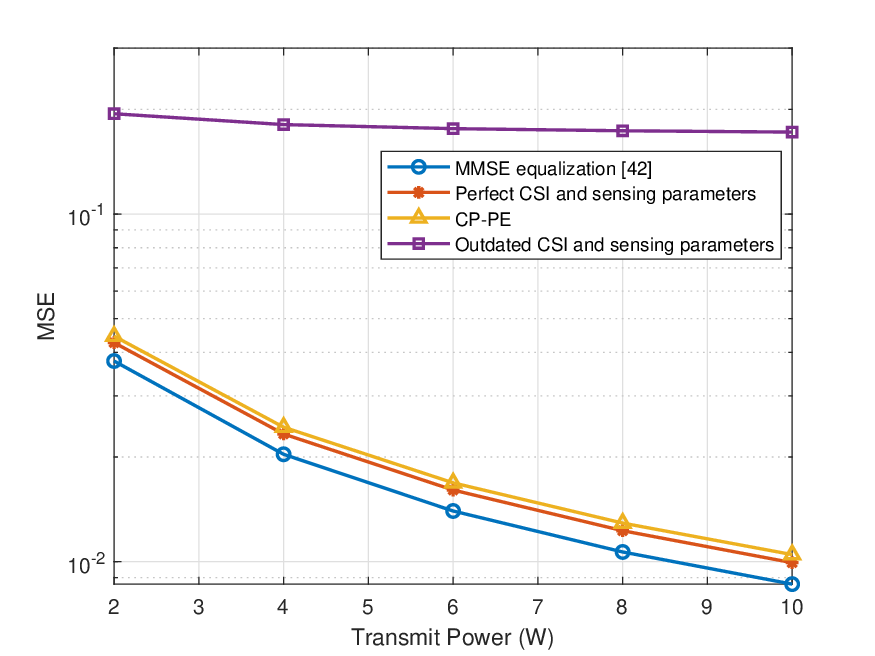}
\caption{Communication MSE performance under different transmit power ($\rho_{C}=1$).}
\label{fig:MSE_C5S0}
\end{figure}

We then investigate the effect of transmit power on the optimized performance of communication and sensing. Fig. \ref{fig:MSE_C5S0} presents the optimized communication MSE versus the transmit power under different schemes, where only the communication performance is considered (i.e., $\rho_C=1$). As illustrated in Fig. \ref{fig:MSE_C5S0}, with increasing transmit power, the MSE of all algorithms shows a decreasing trend. This is because higher transmission power results in a higher SNR for communication symbol demodulation at the UE. Additionally, it can be observed that our proposed algorithm, utilizing the predicted channel for pre-equalization design, not only approaches the communication MSE performance under perfect channel information, indicating the accuracy of our channel prediction algorithm, but also approaches the performance of MMSE equalization based on perfect CSI. This shows the excellent performance of our proposed pre-equalization design scheme in significantly reducing complexity while maintaining significant symbol detection performance. Additionally, the performance curve of pre-equalization designed with outdated CSI shows a significant gap compared to the other three curves. This indicates that pre-equalization designs based on outdated CSI can degrade communication performance, making accurate CSI prediction essential for pre-equalization.

\subsection{Trade-off Between Communication and Sensing Performance}
\begin{figure}[!t]
\centering
\includegraphics[width=\linewidth]{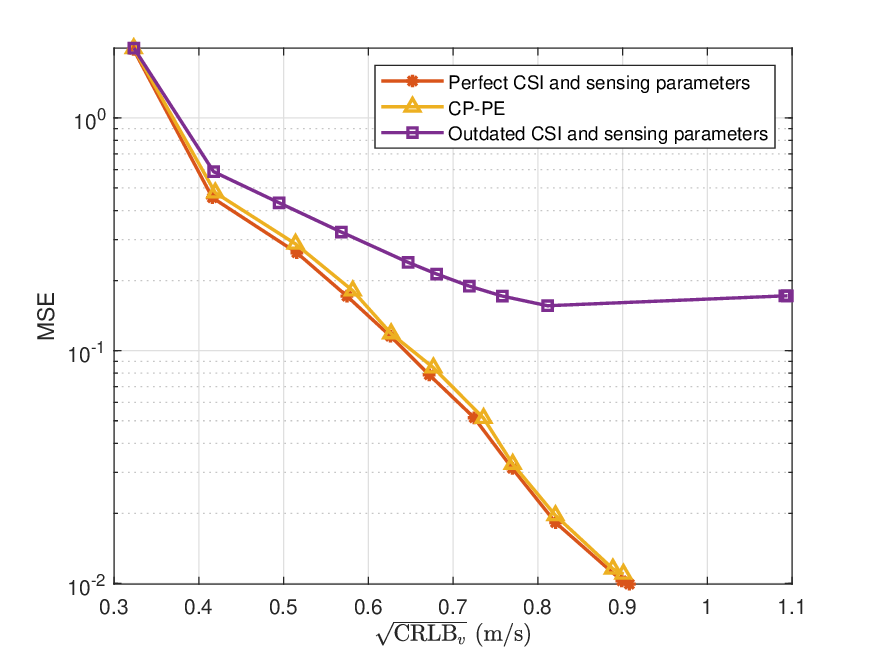}
\caption{Sensing CRLB and communication MSE performance under different weights.}\label{fig:diff_weight}
\end{figure}

Fig. \ref{fig:diff_weight} illustrates the trade-off between communication and sensing performance achieved by different algorithms. As seen in the figure, when the communication weighting factor $\rho_C$ is large, the proposed algorithm achieves an MSE of the order of $10^{-2}$. As the communication weighting factor $\rho_C$ decreases, the sensing CRLB of the CP-PE algorithm is reduced by more than 60\%, indicating a significant improvement in sensing accuracy. This demonstrates that the designed weighting factor allows different pre-equalization designs to adapt to varying communication and sensing requirements.
Moreover, the trade-off curve of the proposed CP-PE algorithm nearly overlaps with that of the design based on perfect CSI and sensing parameters, while the trade-off curve of the pre-equalization design based on outdated CSI always lies outside the trade-off curve of the proposed CP-PE algorithm. This means that, under any communication and sensing requirements, the proposed method can overcome the impact of outdated CSI, achieving a good balance between communication and sensing performance. Note that the CRLB performance of the pre-equalization design based on outdated CSI improves overall with the increase in communication symbol demodulation MSE. However, when the communication weighting factor $\rho_C$ is very large, the MSE increases. This is because when the gap between outdated and perfect CSI and sensing parameters is larger, the better the optimization result based on the outdated CSI, the poorer communication and sensing performance when applied with real CSI. This again underscores the necessity of obtaining accurate CSI.

\begin{figure}[t]
\centering
        \includegraphics[width=\linewidth]{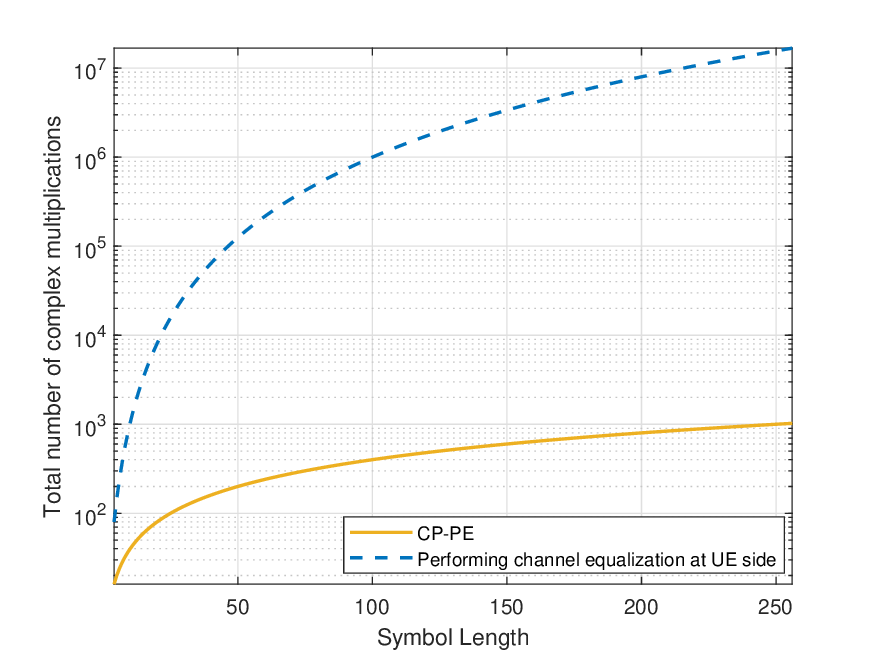}
        \caption{Signal processing complexity at UE side under different symbol lengths.}
        \label{fig:energy}
\end{figure}
\subsection{Communication Receiver Complexity Comparison}
In traditional communication frameworks as shown in Fig. \ref{fig:process}, the UAV UE needs to perform channel equalization before symbol detection, which requires matrix inversion operations. This results in high computational complexity, i.e., $O((MN)^3 + 2^o MN)$, where $o$ is the modulation order. In contrast, the proposed CP-PE framework only requires a complexity of $O(2^o MN)$ at the UE. The comparison of the signal processing complexity, represented by the total number of complex multiplications, between the proposed framework and the traditional framework (i.e., performing channel equalization at the UE side) is plotted in Fig.~\ref{fig:energy}. As shown in the figure, the processing complexity increases with the increase in symbol length. Moreover, the complexity gap between the traditional communication scheme and the proposed CP-PE framework widens significantly. Even for short symbol lengths, the processing complexity is reduced by more than 75\% compared to channel equalization at the UE side. It is worth noting that, given that energy consumption is proportional to complexity, the proposed method greatly reduces energy consumption at the UE, thereby enhancing the battery life of the UAV.

\vspace{-0mm}\section{Conclusions}\label{sec:concl}
In this work, we proposed a pre-equalization design based on the OTFS waveform for an ISAC-enabled air-ground network. To overcome the influence of outdated CSI and sensing parameters, relying on deep learning, a CP-PE framework was developed to effectively design a more precise pre-equalization matrix. The pre-equalization design enabled the receiver to require only simple power normalization for symbol detection, significantly reducing the overhead and complexity associated with channel estimation and equalization at the UE. Simulation results showed that the CP-PE framework enhances both communication and sensing performance compared to pre-equalization based on outdated CSI. Moreover, it closely approximated the performance achieved with perfect CSI and sensing parameters, demonstrating its robustness and suitability for high-mobility scenarios.

 \vspace{-0mm}\bibstyle{IEEEtran}
   \bibliography{IEEEabrv,OTFS_precoding_short}

\end{document}